# Poisson Noise Channel with Dark Current: Numerical Computation of the Optimal Input Distribution

Luca Barletta*, Alex Dytso**
* Politecnico di Milano, Milano, 20133, Italy. Email: luca.barletta@polimi.it
** New Jersey Institute of Technology, Newark, NJ 07102, USA. Email: alex.dytso@njit.edu

**Abstract—This paper considers a discrete time-Poisson noise channel which is used to model pulse-amplitude modulated optical communication with a direct-detection receiver. The goal of this paper is to obtain insights into the capacity and the structure of the capacity-achieving distribution for the channel under the amplitude constraint $\mathsf{A}$ and in the presence of dark current $\lambda$. Using recent theoretical progress on the structure of the capacity-achieving distribution, this paper develops a numerical algorithm, based on the gradient ascent and Blahut-Arimoto algorithms, for computing the capacity and the capacity-achieving distribution. The algorithm is used to perform extensive numerical simulations for various regimes of $\mathsf{A}$ and $\lambda$.**

## I. Introduction

Poisson models are an important set of models with a wide range of applications. In this work, we consider a discrete-time memoryless Poisson channel. For this channel, the input takes values on the set of nonnegative real numbers $\mathbb{R}_0^+$ and the output takes values on the set of nonnegative integers $\mathbb{N}_0$ with the channel law given by

$$P_{Y|X}(k|x) = \frac{1}{k!}(x+\lambda)^k \mathrm{e}^{-(x+\lambda)}, \quad x \in \mathbb{R}_0^+, \, k \in \mathbb{N}_0, \quad (1)$$

where $\lambda \geq 0$ is the nonnegative constant called the *dark current*. In (1) we use the conventions that $0^0 = 1$ and $0! = 1$. That is, (1) models a scenario where, conditioned to $X = x$, the output $Y$ is the Poisson random variable with mean $x + \lambda$.

The objective of this paper is to study the capacity and the capacity-achieving distribution of this channel under the amplitude constraint $\mathsf{A} \geq 0$, that is

$$C(\mathsf{A}, \lambda) = \max_{P_X :\, 0 \leq X \leq \mathsf{A}} I(X; Y), \quad (2)$$

where $P_X$ denotes the distribution of the input random variable $X$, and where $I(X;Y)$ is the mutual information between $X$ and $Y$. We denote by $P_{X^\star}$ the capacity-achieving distribution in (2). The $P_{X^\star}$ and $C(\mathsf{A}, \lambda)$, besides a few cases, are in general unknown. The expression of the capacity is an important benchmark in communications theory, and knowing the capacity-achieving distribution is important as it is useful for the code or modulation design. Furthermore, the availability of numerical examples of $P_{X^\star}$ may also guide theoretical work by pointing out possible properties of $P_{X^\star}$ that need to be proven. Therefore, it is both of practical and theoretical interest to produce numerical example of $P_{X^\star}$ and $C(\mathsf{A}, \lambda)$.

In this work, we take a numerical approach to computing $P_{X^\star}$ and $C(\mathsf{A}, \lambda)$ and produce extensive numerical examples of these quantities for various parameter regimes of $\mathsf{A}$ and $\lambda$. Data from these simulations are made available at [1].

Our ability to produce extensive numerical example of $P_{X^\star}$ is enabled by a recent result in [2], which produced a first firm upper bound of the order $\mathsf{A} \log^2(\mathsf{A})$ on the size of the support of $P_{X^\star}$. A firm upper bound on the size of the support in [2], allows us to move the optimization from the space of probability distributions to the $\mathbb{R}^{2n}$ space where $n$ is the number of support points of $P_{X^\star}$. Working in $\mathbb{R}^{2n}$ allows us to employ finite-dimensional methods such as the gradient ascent together with the Blahut-Arimoto algorithm [3]. The firm bound $n$ also guarantees that our algorithms will converge to at least a local maximum. To guarantee our solution is close to the global maximum, we also check it against the Karush-Kuhn-Tucker (KKT) conditions. The details of the implementation are provided in Section II.

### A. Outline and Contributions

The outline and the contribution of the paper are as follows. The remaining part of Section I will present our notation and will survey the relevant literature. Section II will present the following: the KKT conditions needed for the optimality of $P_{X^\star}$; the algorithm that will produce numerical examples of $P_{X^\star}$; and the discussion of the objectives we seek to achieve with the numerical simulations. Section III will present our extensive numerical simulations and the observations surrounding these simulations.

### B. Notation

Throughout the paper, the deterministic scalar quantities are denoted by lower-case letters and random variables are denoted by uppercase letters.

We denote the distribution of a random variable $X$ by $P_X$. The support set of $P_X$ is denoted and defined as

$$\mathsf{supp}(P_X) = \{ x : \text{for every open set } \mathcal{D} \ni x \text{ we have that } P_X(\mathcal{D}) > 0 \}. \quad (3)$$

The cardinality of $\mathsf{supp}(P_X)$ will be denoted by $|\mathsf{supp}(P_X)|$. The relative entropy between distributions $P$ and $Q$ will be denoted by $\mathsf{D}(P\|Q)$.

Let $g_1$ and $g_2$ be nonnegative functions, then

- $g_1(x) = O(g_2(x))$ means that there exists a constant $c > 0$ and $x_0$ such that $\frac{g_1(x)}{g_2(x)} \leq c$ for all $x > x_0$;
- $g_1(x) = \Omega(g_2(x))$ means that $g_2(x) = O(g_1(x))$;
- $g_1(x) = \Theta(g_2(n))$ means that $g_1(x) = O(g_2(x))$ and $g_1(x) = \Omega(g_2(x))$; and
- $g_1(x) = o(g_2(x))$ means $\lim_{x\to\infty} \frac{g_1(x)}{g_2(x)} = 0$.

### C. Past Work

The discrete-time Poisson channel is suited to model low-intensity, direct-detection optical communication channels [4].

The first study of capacity and of the capacity-achieving distribution for the Poisson channel was done in [5], where the author consider the case of zero dark-current (i.e., $\lambda = 0$) in the presence of both amplitude and average-power constraints. The authors of [5] showed that the optimal input distribution for any $\mathsf{A}$ can contain at most one mass point between $(0, 1)$.

In [6], the author studied the KKT conditions for the capacity-achieving distribution, which led to the conclusion that distribution is unique and discrete with finitely many mass points. Moreover, in [6], for the case of zero dark current, it was shown that binary communication is optimal if and only if $\mathsf{A} \leq \bar{\mathsf{A}}$ where $\bar{\mathsf{A}} \approx 3.3679$. Further, studies of the binary communication in presence of nonzero dark current and/or additional average-power constraints where undertaken in [7] where sufficient and necessary conditions for optimality have been produced. The study in [7] has also shown that the support of the optimal input distribution always contains points 0 and $\mathsf{A}$.

For the case of $\lambda = 0$, the authors of [2] provided several new properties of the optimal input distribution as well as a new compact characterization of the capacity. For example, in [2], it has been shown that the support of $P_{X^\star}$ satisfies the following:

$$\Omega(\sqrt{\mathsf{A}}) \leq |\mathsf{supp}(P_{X^\star})| \leq O\left(\mathsf{A} \log^2(\mathsf{A})\right). \tag{4}$$

As already was alluded to previously, these bounds enable us to run finite-dimensional optimization algorithms. We note that the upper bound in (4) holds even if $\lambda > 0$ since increasing the dark current cannot increase the size of the support.

Numerical examples of the capacity-achieving distributions have also been provided in [8] and [7] for the channel with both amplitude and average-power constraints. In contrast, this paper focuses only on the amplitude constraint and provides more extensive simulations for this setting. Moreover, our algorithm has convergence grantees due to the finite-dimensional nature of our algorithms. We note that there are several other numerical recipes for generating an optimal input distribution, and the interested reader is referred to [3], [9], [10]. However, most of these approaches ultimately optimize over the space of distributions, which is an infinite-dimensional space.

The capacity-achieving distribution with only an average-power constraint was considered in [11] and was shown to be discrete with infinitely many mass points. The low-average-power and the low-amplitude asymptotics of the capacity have been studied in [12]–[16]. A number of papers have also focused on upper and lower bounds on the capacity. The first upper and lower bounds on the capacity have been derived in [5] for two situations: the case of the average-power constraint only, and the case of both the average-power and the amplitude constraint with $\mathsf{A} \leq 1$. The authors of [17] derived upper and lower bounds, in the case of the average-power constraint only, by focusing on the regime where both power and the dark current tend to infinity with a fixed ratio. Firm upper and lower bounds on the capacity in the case of only the average-power constraint and no dark current have been derived in [12] and [13]. Bounds in [12] and [13] have been further improved in [11] and [18]. The most general bounds on the capacity that consider both the amplitude and the average-power constraints on the input and hold for an arbitrary value of the dark current have been derived in [19]. The bounds in [19] have been shown to be tight in the regime where both the average-power and the amplitude constraint approach infinity with a fixed ratio of amplitude to average-power. Finally, the authors of [20] sharpened the results of [19].

## II. Proposed Algorithm

In this section, we first present the KKT equations, which provide sufficient and necessary conditions for the optimality of $P_{X^\star}$. Second, we present the algorithm which will be used to generate numerical examples of $P_{X^\star}$. The proposed algorithm is the combination of the Blahut-Arimoto algorithm and the gradient ascent algorithm. The key step of the algorithm evaluates the KKT conditions with the candidate solution, which ensures that the solution is close to the true solution. The final part of this section discusses some of the objectives that we aim to achieve with our numerical simulations.

### A. KKT conditions

The starting point for the design of our algorithm are the following KKT conditions shown in [6].

**Lemma 1.** *The capacity-achieving distribution $P_{X^\star}$ and induced capacity-achieving output distribution $P_{Y^\star}$ satisfy the following:*

$$i(x; P_{X^\star}) \leq C(\mathsf{A}, \lambda), \quad x \in [0, \mathsf{A}] \tag{5a}$$

$$i(x; P_{X^\star}) = C(\mathsf{A}, \lambda), \quad x \in \mathsf{supp}(P_{X^\star}), \tag{5b}$$

*where*

$$i(x; P_{X^\star}) = \mathsf{D}(P_{Y|X}(\cdot|x) \| P_{Y^\star}). \tag{6}$$

In the numerical computation of $P_X$ we find more convenient to check the negated version of (5), where a tolerance parameter $\varepsilon$ is introduced which trades off accuracy with computational burden. Specifically, $P_X$ is not an optimal input pmf if any of the following conditions is satisfied:

$$i(0; P_X) + \varepsilon < i(x; P_X), \text{ for some } x \in [0, \mathsf{A}] \tag{7a}$$

$$|i(x;P_X) - i(0;P_X)| > \varepsilon, \text{ for some } x \in \mathsf{supp}(P_X). \quad \text{(7b)}$$

Note that in (7) in place of the capacity $C(\mathsf{A},\lambda)$, which is unknown, we used the value of $i(0;P_X)$, thanks to the fact that $0 \in \mathsf{supp}(P_{X^\star})$ for any $\lambda$. With some abuse of notation, we refer to (7) as to the $\varepsilon$-KKT conditions.

### B. Numerical Algorithm

The algorithm that gives a numerical lower bound to capacity and that estimates the optimal input pmf $P_{X^\star}$ is given in Algorithm 1. The algorithm takes as input: the value $\mathsf{A}$ of the amplitude constraint; the vector $\mathbf{x}$ of the support points of the initial tentative $P_X$; the vector $\mathbf{p}$ of the probabilities associated with the points in $\mathbf{x}$; and a value $\varepsilon$ which is related to the accuracy of the capacity estimation. Using the bound in [2], we can set the dimensionally of $\mathbf{p}$ and $\mathbf{x}$ to be less than $\mathsf{A}\log^2(\mathsf{A})$.

The optimization procedure is iterative and divided into two parts: (i) $N_{\mathsf{BA}}$ iterations of the classic Blahut-Arimoto algorithm are used to let the tentative pmf converge to stable values, and (ii) $N_{\mathsf{GA}}$ iterations of a gradient ascent algorithm are used to modify the positions of the support points of the tentative input distribution. The objective function of the gradient ascent is the mutual information. We use a backtracking line search version of the gradient ascent [21], in order to ensure that the sequence of mutual information values obtained along the iterations is nondecreasing, which ensures convergence to a local maximum.

After the two-step optimization, if there is a set of support points that are driven too close to each other by the gradient ascent, then such points are clustered together in a new support point which assumes a probability equal to the sum of the probabilities of the clustered points. Note that the clustering operation is equivalent to having a minimum distance constraint among support points. We have observed that this constraint in general helps numerical stability. We have chosen a minimum distance of $10^{-2}$, which seems to be an inactive capacity constraint for all values of $\mathsf{A} > 10^{-2}$ and of $\lambda \geq 0$. We caution, however, this constraint might be active around the transition points (i.e., when one point splits into two), but becomes inactive once again after we move away by $10^{-2}$ from the transition point.

The optimality of the resulting distribution $P_X$ is tested with the $\varepsilon$-KKT conditions. If any of the conditions (7) is satisfied, then the function UPDATE modifies $P_X$ according to the following rules. Let

$$\widehat{x} = \arg\max_{x\in[0,\mathsf{A}]} i(x;P_X) \quad (8)$$

be the candidate novel support point, and

$$\mathcal{S} = \{x \in \mathsf{supp}(P_X) : \text{(7b) is true}\} \quad (9)$$

be the set of support points for which the information density falls outside the $\varepsilon$-strip that contains $i(0;P_X)$. If both conditions in (7) are verified and there exist $x_1, x_2 \in \mathcal{S}$ such that $|x_1 - x_2| < \delta = 0.1$ and $\widehat{x} \in [x_1, x_2]$, then $\widehat{x}$ replaces both $x_1$ and $x_2$ in the support of $P_X$ with $P_X(\widehat{x}) = P_X(x_1) + P_X(x_2)$. Otherwise, if only (7a) is satisfied, then $\widehat{x}$ is added to the support of $P_X$ and the probabilities are set to $P_X(x) = |\mathsf{supp}(P_X)|^{-1}$ for all $x \in \mathsf{supp}(P_X)$.

**Algorithm 1** Capacity and input PMF estimation

```
procedure MAIN(A, x, p, ε)
    N_BA ← 100 / Number of Blahut-Arimoto iterations
    N_GA ← 20 / Number of gradient ascent iterations
    repeat
        k ← 0
        while k < 100 do
            k ← k + 1
            p ← BLAHUT-ARIMOTO(x, p, N_BA)
            x ← GRADIENT-ASCENT(x, p, N_GA)
        end while
        (x, p) ← CLUSTER(x, p)
        valid ← KKT-VALIDATION(x, p, ε)
        if valid = False then
            (x, p) ← UPDATE(x, p)
        end if
    until valid = True
    C(A) ← I(x, p)
    return x, p, C(A)
end procedure
```

The main algorithm repeats the two-step optimization procedure until $P_X$ is validated, which becomes the proposed $P_{X^\star}$.

### C. Our Objectives

The first objective of this work is to produce numerical examples of $C(\mathsf{A},\lambda)$ for various parameter regimes of $(\mathsf{A},\lambda)$. The second objective is to study properties of $P_{X^\star}$. In terms of properties, we are the most interested in the structure of the support of $P_{X^\star}$. More specifically, we are interested in the locations of the support points and the cardinality of the support. As was already mentioned in Section I, the size of the support is upper bounded by $\mathsf{A}\log^2(\mathsf{A})$. The lower bound on the support can be provided by using the bound in [2, Rem. 3]:

$$|\mathsf{supp}(P_{X^\star})| \geq \mathrm{e}^{C(\mathsf{A},\lambda)}. \quad (10)$$

However, since $C(\mathsf{A},\lambda)$ is not available, it seems like the bound in (10) is not computable. Note, however that from the definition of capacity we have that $C(\mathsf{A},\lambda) \geq I(\widetilde{X};\widetilde{Y})$ for any random variable $\widetilde{X} \in [0,\mathsf{A}]$ and where $\widetilde{Y}$ is induced by $\widetilde{X}$. Therefore, since we can choose any $\widetilde{X} \in [0,\mathsf{A}]$, we selected it to be the one that is the output of the numerical simulation, and following lower bound holds:

$$|\mathsf{supp}(P_{X^\star})| \geq \mathrm{e}^{I(\widetilde{X};\widetilde{Y})}. \quad (11)$$

## III. Simulations and Observations

In this section, we present the results of the numerical simulations performed by Algorithm 1. We set the tolerance value to be $\varepsilon = 10^{-6}$. For several values of the parameters $(\mathsf{A},\lambda)$, we will plot the following: $C(\mathsf{A},\lambda)$; the location of the support points of $P_{X^\star}$; the cardinality of $\mathsf{supp}(P_{X^\star})$; and the lower bound on the cardinality of the support in (11). The simulation data results are publicly available [1].

### A. Zero Dark Current

In this section, we study the structure of the capacity-achieving distribution when the value of the dark current is zero (i.e., $\lambda = 0$). For the most up to date theoretical results on the structure of the capacity-achieving distribution in the case when $\lambda = 0$ the interested reader is referred to [2]. In [2], the input distributions were simulated for values $\mathsf{A} \leq 15$. In this section, we produce optimal input distributions for values of A up to 128.

Fig. 1 presents the structure of the support of the capacity-achieving distribution vs. A and Fig. 6a shows $C(\mathsf{A}, 0)$ vs. A. From these figures we observe the following:

- The number of points as A grows increases by at most one - a point in the middle splits into at most two points. Therefore, an interesting theoretical future direction would be to prove that this is indeed the case. From a practical point of view, proving this conjecture can further reduce the computational complexity of numerical solutions. We note, however, that proving this might be a challenging task. For instance, a similar conjecture has been proposed for the Gaussian noise channel in [22] and, as of the writing of this paper, has not yet been resolved.
- For $\mathsf{A} > 1$, the second smallest point is larger than one and asymptotically appears to converge to the value of 2.4. We note that already in [5] the authors showed that the optimal input distribution for any A contains at most one mass point in the open interval $(0, 1)$. Therefore, an interesting future direction is to strengthen this statement and show that for $\mathsf{A} > 1$, there exists no mass point in the open interval $(0, 1)$.
- The distance between the second-largest point and A appears to be growing. A possible future direction can consist of showing this. In fact, already in [2, Thm. 1], it has been shown that the distance between A and the second-largest point is at least one. Finally, Fig. 1b suggests that, if the distance does grow, the rate of growth must be $o(\mathsf{A})$.
- The numerical simulations suggest that the cardinality of $\mathsf{supp}(P_{X^\star})$ behaves as $\Theta(\sqrt{\mathsf{A}})$. Note that it has been shown in [2] that the lower bound in (10) also has a $\sqrt{\mathsf{A}}$ behavior and was conjectured to be the correct order. Thus these simulation results seem to confirm that conjecture. Finally, note that the simulation results of the lower bound in (11), in Fig. 1c, however, seem somewhat off. We speculate that the reason for this might be a very small constant in front of the $\sqrt{\mathsf{A}}$ term, which does not change the order but does impact the shape of the bound.

### B. Nonzero Dark Current

In this section, we study the structure of the capacity-achieving distribution when the value of the dark current is greater than zero.

Fig. 2, Fig. 3, and Fig. 4 depict $\mathsf{supp}(P_{X^\star})$ vs. A for $\lambda = 1$, $\lambda = 10$ and $\lambda = 100$, respectively. Fig. 5 shows $\mathsf{supp}(P_{X^\star})$ vs. $\lambda$ for $\mathsf{A} = 50$. The corresponding capacity values are plotted in Fig. 6. While many of the general observations about the structure of $\mathsf{supp}(P_{X^\star})$ appear to be the same for $\lambda = 0$ and $\lambda > 0$, there are some key differences. From these figures we observe the following.

- The positions of the second smallest points appear to be heavily dependent on $\lambda$. For example, for $\lambda = 1$ the second smallest point is always larger than two and appears to asymptotically approach a value of 4.6. As a future direction, it would be interesting to find a lower bound on the location of the second smallest point as a function of $\lambda$.
- The cardinality of $\mathsf{supp}(P_{X^\star})$ as a function of A also appears to have $\Theta(\sqrt{\mathsf{A}})$ behavior albeit with a factor that depend on $\lambda$. Indeed, by using a lower bound on the capacity in [19, Thm. 4] together with the lower bound in (10) we have that
$$|\mathsf{supp}(P_{X^\star})| \geq c_1 \sqrt{\mathsf{A}} \mathrm{e}^{-c_2 \sqrt{\frac{\lambda}{\mathsf{A}}}}, \tag{12}$$
for some positive constants $c_1, c_2$. Therefore, the simulation suggests that the lower bound in (12) is tight.
- For large values of dark current, the Poisson channel with input $X = x$ begins to behave as a Gaussian channel with mean $x + \lambda$ and variance $x + \lambda$; the interested reader is referred to [23], [24] where the capacity of this and other approximations have been considered. The large $\lambda$ behavior of the capacity-achieving distribution is depicted in Fig. 4 and Fig. 5. As one possible future direction, it would be interesting to characterize and understand the structure of the capacity-achieving distribution for these Gaussian models and then compare the structure of the capacity-achieving distributions to those for the Poisson noise channel in Fig. 4 and Fig. 5. As an even more ambitious direction, one could study whether the Gaussian approximation's capacity-achieving distributions converge to the Poisson channel's capacity-achieving distributions in some distance over a probability space (e.g., Lévy metric).

## References

[1] L. Barletta and A. Dytso. (2021) Simulated data. [Online]. Available: https://github.com/ucando83/PoissonCapacity

[2] A. Dytso, L. Barletta, and S. Shamai, "Properties of the support of the capacity-achieving distribution of the amplitude-constrained poisson noise channel," *IEEE Transactions on Information Theory*, 2021.

[3] R. Blahut, "Computation of channel capacity and rate-distortion functions," *IEEE Transactions on Information Theory*, vol. 18, no. 4, pp. 460–473, 1972.

[4] J. P. Gordon, "Quantum effects in communications systems," *Proceedings of the IRE*, vol. 50, no. 9, pp. 1898–1908, 1962.

[5] R. McEliece, E. Rodemich, and A. Rubin, "The practical limits of photon communication," *Jet Propulsion Laboratory Deep Space Network Progress Reports*, vol. 42, pp. 63–67, 1979.

[6] S. S. Shamai, "Capacity of a pulse amplitude modulated direct detection photon channel," *IEE Proceedings I (Communications, Speech and Vision)*, vol. 137, no. 6, pp. 424–430, 1990.

[7] J. Cao, S. Hranilovic, and J. Chen, "Capacity-achieving distributions for the discrete-time Poisson channel Part ii: Binary inputs," *IEEE Transactions on Communications*, vol. 62, no. 1, pp. 203–213, 2014.

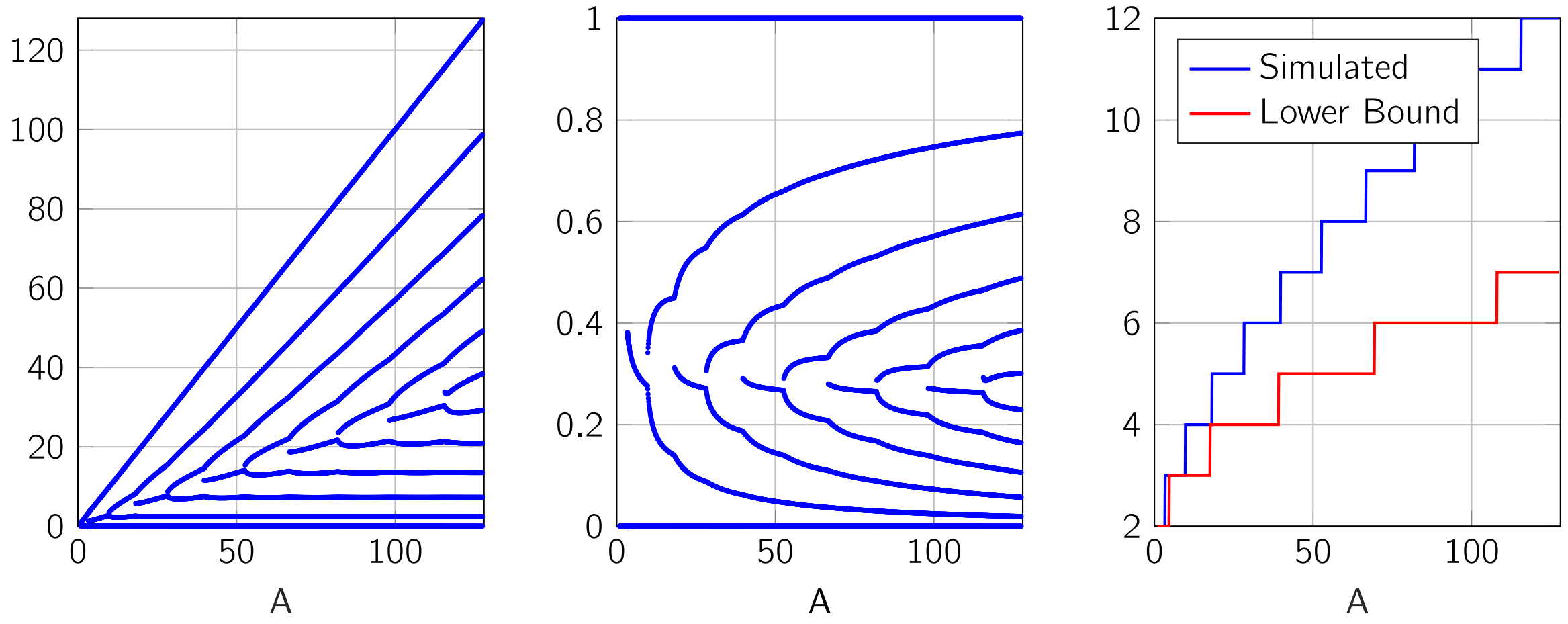

(a) Plot of $\textbf{supp}(P_{X^\star})$ vs. A. (b) Plot of $\frac{\text{supp}(P_{X^\star})}{A}$ vs. A. (c) Plot of $|\textbf{supp}(P_{X^\star})|$ vs. A.

Fig. 1: Simulations Results for the case of $\lambda = 0$.

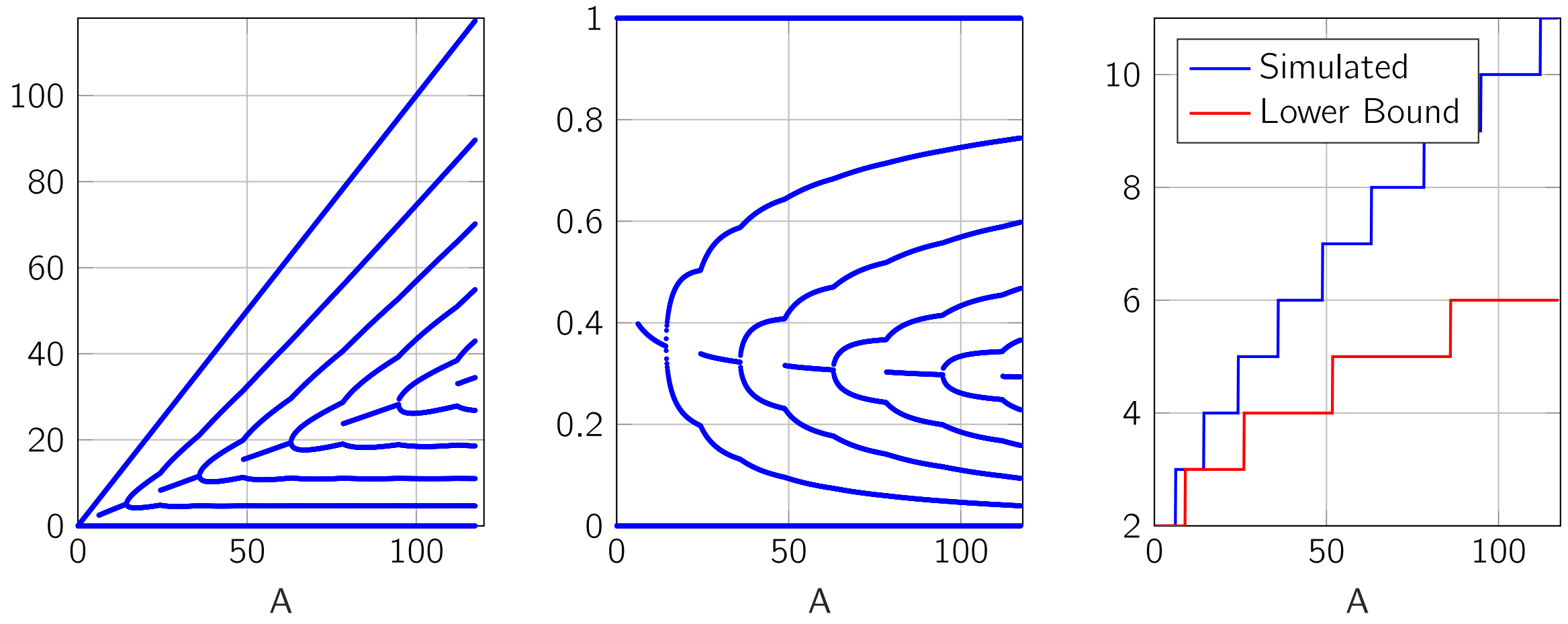

(a) Plot of $\textbf{supp}(P_{X^\star})$ vs. A. (b) Plot of $\frac{\text{supp}(P_{X^\star})}{A}$ vs. A. (c) Plot of $|\textbf{supp}(P_{X^\star})|$ vs. A.

Fig. 2: Simulations results for the case of $\lambda = 1$.

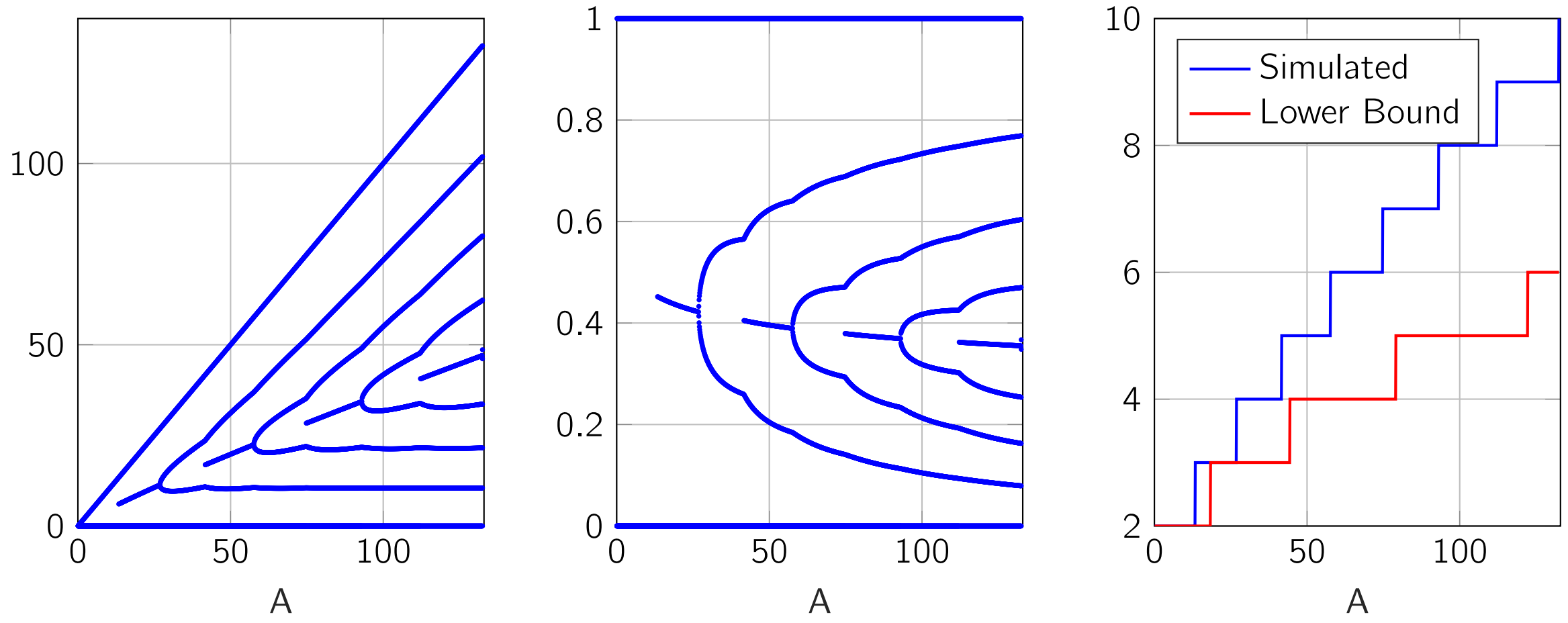

(a) Plot of $\textbf{supp}(P_{X^\star})$ vs. A. (b) Plot of $\frac{\text{supp}(P_{X^\star})}{A}$ vs. A. (c) Plot of $|\textbf{supp}(P_{X^\star})|$ vs. A.

Fig. 3: Simulations results for the case of $\lambda = 10$.

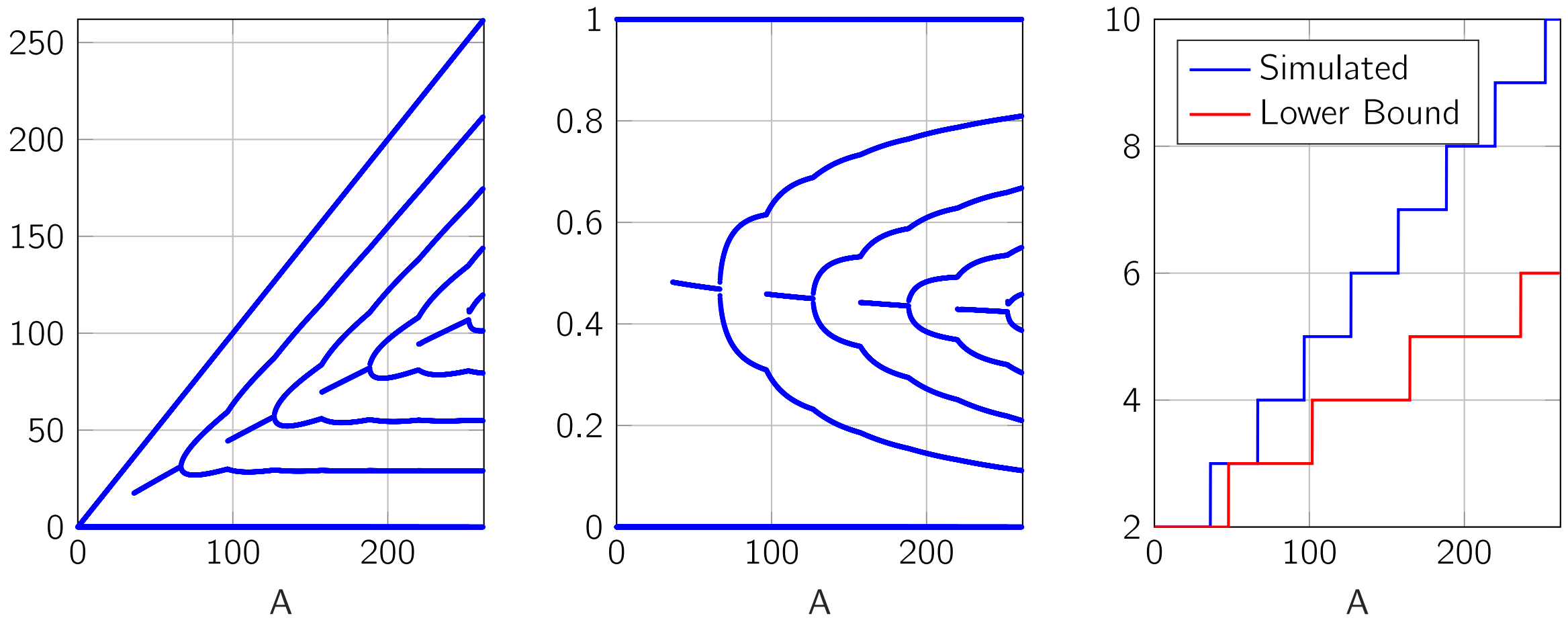

(a) Plot of $\mathbf{supp}(P_{X^\star})$ vs. A. (b) Plot of $\frac{\mathbf{supp}(P_{X^\star})}{\mathsf{A}}$ vs. A. (c) Plot of $|\mathbf{supp}(P_{X^\star})|$ vs. A.

Fig. 4: Simulations results for the case of $\lambda = 100$.

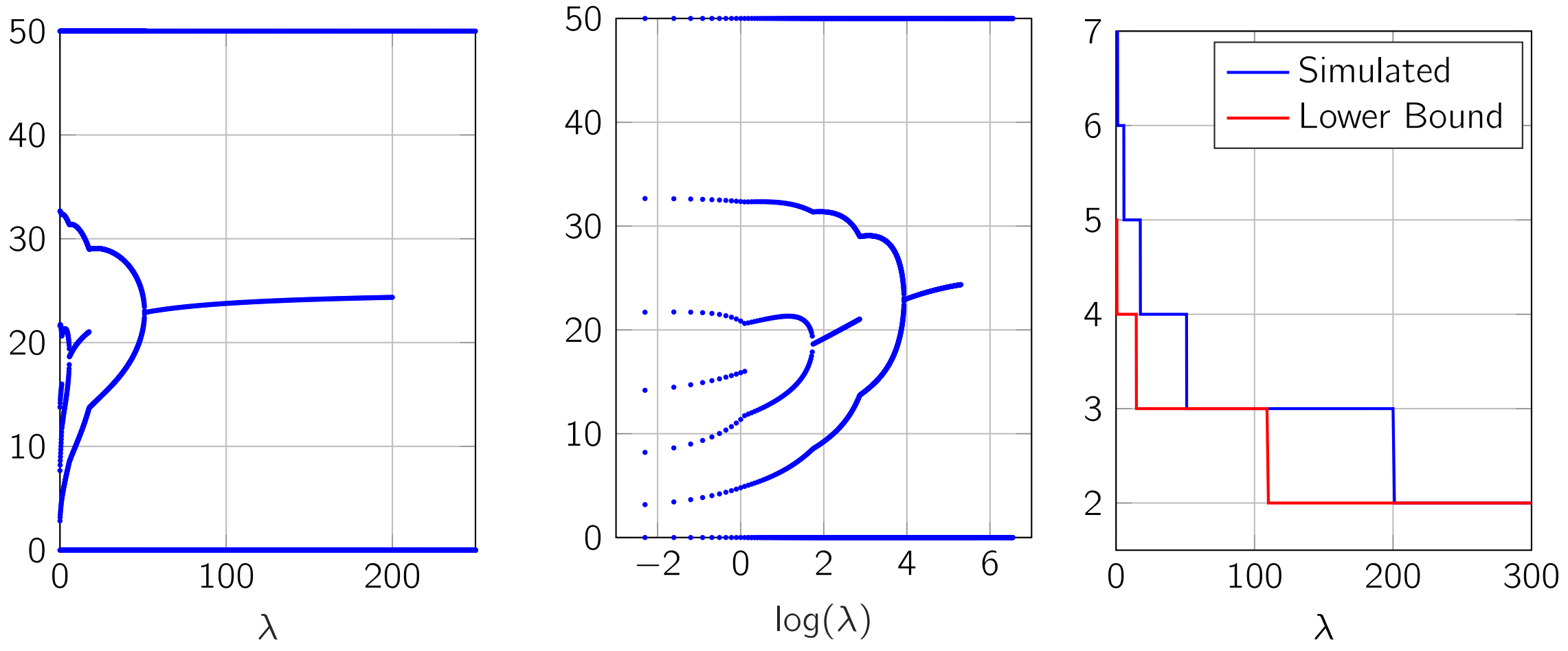

(a) Plot of $\mathbf{supp}(P_{X^\star})$ vs. $\lambda$. (b) Plot of $\mathbf{supp}(P_{X^\star})$ vs. $\log(\lambda)$. (c) Plot of $|\mathbf{supp}(P_{X^\star})|$ vs. A.

Fig. 5: Simulations results for the case of $\mathsf{A} = 50$.

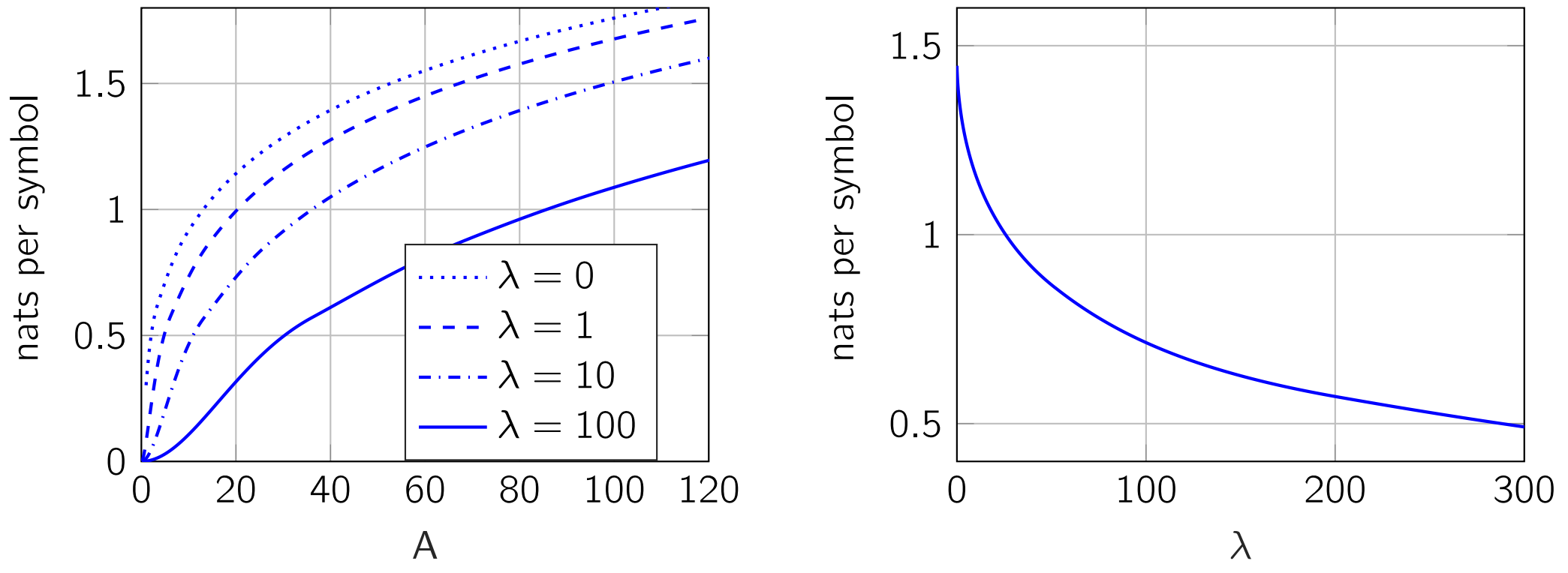

(a) Plot of $C(\mathsf{A}, \lambda)$ vs. A for $\lambda = 0, 1, 10$ and $\lambda = 100$. (b) Plot of $C(\mathsf{A}, \lambda)$ vs. $\lambda$ for $\mathsf{A} = 50$.

Fig. 6: Capacity plots.

[8] ——, "Capacity-achieving distributions for the discrete-time Poisson channel Part i: General properties and numerical techniques," *IEEE Transactions on Communications*, vol. 62, no. 1, pp. 194–202, 2014.

[9] C.-I. Chang and L. D. Davisson, "On calculating the capacity of an infinite-input finite (infinite)-output channel," *IEEE Transactions on Information Theory*, vol. 34, no. 5, pp. 1004–1010, 1988.

[10] J. Huang and S. P. Meyn, "Characterization and computation of optimal distributions for channel coding," *IEEE Transactions on Information Theory*, vol. 51, no. 7, pp. 2336–2351, 2005.

[11] M. Cheraghchi and J. Ribeiro, "Improved upper bounds and structural results on the capacity of the discrete-time Poisson channel," *IEEE Transactions on Information Theory*, vol. 65, no. 7, pp. 4052–4068, 2019.

[12] A. Martinez, "Spectral efficiency of optical direct detection," *JOSA B*, vol. 24, no. 4, pp. 739–749, 2007.

[13] ——, "Achievability of the rate $\frac{1}{2}\log(1+\epsilon_s)$ in the discrete-time Poisson channel," *arXiv preprint arXiv:0809.3370*, 2008.

[14] A. Lapidoth, J. H. Shapiro, V. Venkatesan, and L. Wang, "The discrete-time Poisson channel at low input powers," *IEEE Transactions on Information Theory*, vol. 57, no. 6, pp. 3260–3272, 2011.

[15] L. Wang and G. W. Wornell, "A refined analysis of the Poisson channel in the high-photon-efficiency regime," *IEEE Transactions on Information Theory*, vol. 60, no. 7, pp. 4299–4311, 2014.

[16] ——, "The impact of dark current on the wideband Poisson channel," in *2014 IEEE International Symposium on Information Theory (ISIT)*. IEEE, 2014, pp. 2924–2928.

[17] D. Brady and S. Verdú, "The asymptotic capacity of the direct detection photon channel with a bandwidth constraint," in *28th Allerton Conf. Commun., Control and Comp.*, Oct. 1990, pp. 691–700.

[18] M. Cheraghchi and J. Ribeiro, "Non-asymptotic capacity upper bounds for the discrete-time Poisson channel with positive dark current," *arXiv preprint arXiv:2010.14858*, 2020.

[19] A. Lapidoth and S. M. Moser, "On the capacity of the discrete-time Poisson channel," *IEEE Transactions on Information Theory*, vol. 55, no. 1, pp. 303–322, 2009.

[20] Y. Yu, Z. Zhang, L. Wu, and J. Dang, "Lower bounds on the capacity for Poisson optical channel," in *2014 Sixth International Conference on Wireless Communications and Signal Processing (WCSP)*. IEEE, 2014, pp. 1–5.

[21] S. Boyd, S. P. Boyd, and L. Vandenberghe, *Convex Optimization*. Cambridge university press, 2004.

[22] N. Sharma and S. Shamai (Shitz), "Transition points in the capacity-achieving distribution for the peak-power limited AWGN and free-space optical intensity channels," *Probl. Inf. Transm.*, vol. 46, no. 4, pp. 283–299, 2010.

[23] A. Tsiatmas, F. M. Willems, and C. P. Baggen, "Square root approximation to the Poisson channel," in *2013 IEEE International Symposium on Information Theory*. IEEE, 2013, pp. 1695–1699.

[24] S. M. Moser, "Capacity results of an optical intensity channel with input-dependent Gaussian noise," *IEEE Transactions on Information Theory*, vol. 58, no. 1, pp. 207–223, 2012.